\documentclass[epj]{webofc}
\usepackage[varg]{txfonts}   
\wocname{EPJ Web of Conferences}
\woctitle{ICNFP 2015}
\begin{document}
\selectlanguage{english}
\title{A possible evidence of observation of two mixed phases in nuclear collisions}
%
%

\author{K.A. Bugaev\inst{1} 
\and
        A.I. Ivanytskyi\inst{1}
        \and
         V.V. Sagun\inst{1}, G.M. Zinovjev\inst{1} 
         \and
        D.R. Oliinychenko\inst{1,2}
        \and
        V.S. Trubnikov\inst{3}
        \and
        E.G. Nikonov\inst{4}
}

\institute{Bogolyubov Institute for Theoretical Physics of the National Academy of Sciences of Ukraine, 03680, Kiev,  Ukraine
\and
FIAS, Goethe University,  Ruth-Moufang Str. 1, 60438 Frankfurt upon Main, Germany
\and
 National Science Center ``Kharkov Institute of Physics and Technology'', 61108,  Kharkov, Ukraine
 \and
 Laboratory for Information Technologies, JINR, Joliot-Curie Str.  6,   Dubna 141980, Russia
}

\abstract{%
	Using an advanced version of the hadron resonance gas model we have found several remarkable irregularities at chemical freeze-out. The most prominent of them are two sets of highly correlated quasi-plateaus in the collision energy dependence of the entropy per baryon, total pion number per baryon, and thermal pion number per baryon which we found at center of mass  energies 3.6-4.9 GeV and 7.6-10 GeV. The low energy set of quasi-plateaus was predicted a long time ago.  On the basis of the generalized shock-adiabat model we demonstrate that the low energy correlated quasi-plateaus give evidence for the anomalous thermodynamic properties of the mixed phase at its boundary to the quark-gluon plasma. The question is whether the high energy correlated quasi-plateaus are also related to some kind of mixed phase. In order to answer this question we employ  the results of  a systematic meta-analysis of the  quality of  data description of 10 existing event generators of nucleus-nucleus collisions in the range of center of mass collision energies from 3.1 GeV to 17.3 GeV. 
These generators   are divided into two groups: the first group includes the generators which account for the quark-gluon plasma formation  during nuclear collisions, while the second group includes the generators which do not assume the  quark-gluon plasma   formation in such collisions. 
Comparing the  quality  of data description of more than a hundred of different data sets of strange hadrons by these two groups of generators, we find two regions of the equal quality  of data  description which are located at the center of mass collision energies 4.3-4.9 GeV and 10.-13.5 GeV. These two regions of equal quality  of data  description  we interpret as regions  of the  hadron-quark-gluon mixed phase formation. Such a conclusion is strongly supported by the irregularities in the collision energy dependence  of the experimental ratios of the Lambda hyperon number per proton and positive kaon number per Lambda hyperon. 
Although at the moment it is unclear, whether these regions belong to the same mixed phase or not, there are arguments  that the most probable collision energy range to probe the QCD phase diagram (tri)critical endpoint is 12-14 GeV. 
}
\maketitle
\section{Introduction}
\label{intro_Bugaev}

\begin{figure}[h]
\centering
\mbox{\includegraphics[width=70mm,clip]{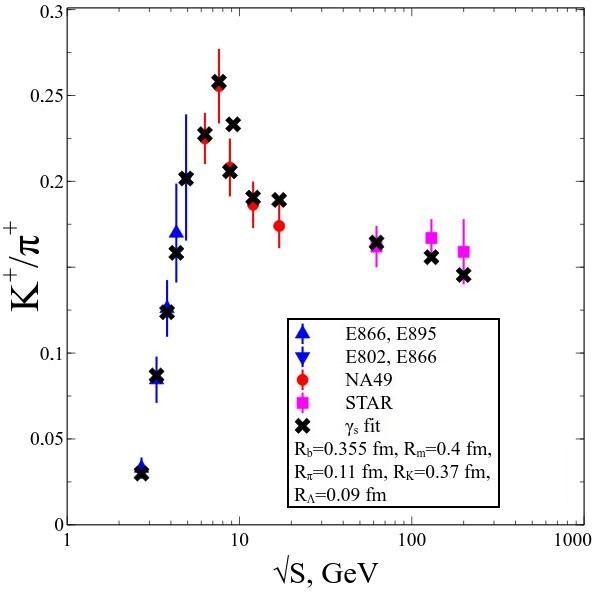}\hspace*{0.2mm}\includegraphics[width=70mm,clip]{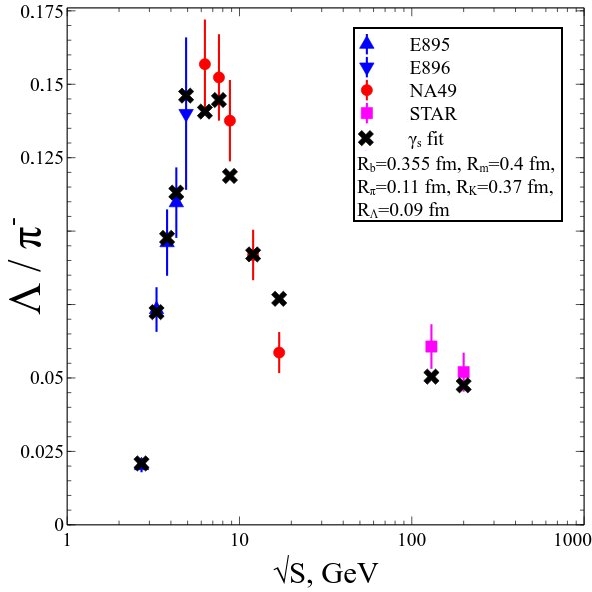}}
\caption{Collision energy dependence of  $K^+/\pi^+$  and $\Lambda/\pi^-$ hadron yield ratios which traditionally were the most problematic ones. }
\label{fig_Bugaev_Horn}       
\end{figure}

The hadron resonance gas model (HRGM) \cite{Andronic:05} is traditionally used to extract the parameters of chemical freeze-out (CFO) from the measured hadronic yields.  Its version with the multicomponent hard-core repulsion
 \cite{Oliinychenko:12,HRGM:13,SFO:13}
allowed one for the first time to successfully describe the most problematic ratios $K^+/\pi^+$ with $\chi^2/dof = 3.9/14$ and 
$\Lambda/\pi^-$ with $\chi^2/dof = 10.2/12$ without spoiling all other  hadron yield ratios \cite{Sagun,Sagun2}. 
Fig.~\ref{fig_Bugaev_Horn}  demonstrates the present  fit quality of  these traditionally problematic ratios. 
 The  achieved high quality $\chi^2/dof \simeq 0.95$  \cite{Sagun,Sagun2} of data description   of 111 independent hadron yield ratios measured at midrapidity  in central nucleus-nucleus collisions at the center of mass energies $\sqrt s_{NN}=2.7, 3.1, 3.6, 4.3, 4.9, 6.3, 7.6,  8.8,  9.2, 12.3, 17.3, 62.4, 130, 200$ GeV
proves that the multicomponent  version of HRGM is a precise and a sensitive tool  of  heavy ion collision phenomenology. 

Using the multicomponent  version of HRGM it was possible to reveal a few  novel  irregularities at CFO.
The most remarkable of them are  two sets of highly correlated quasi-plateaus in the collision energy dependence of the entropy per baryon, total pion number per baryon, and thermal pion number per baryon which were found at the center of mass energies 3.6-4.9 GeV and 7.6-10 GeV \cite{Bugaev:SA1} and  the sharp peak of the trace anomaly found  at the center of mass energy 4.9 GeV \cite{Bugaev:SA2}.  The low energy set of quasi-plateaus was predicted a long time ago \cite{KAB:89a,KAB:90,KAB:91} as a signal of the  anomalous thermodynamic properties inside the quark-gluon-hadron mixed phase. Unfortunately, the generalized shock-adiabat model  cannot be safely applied to the central nuclear collisions 
at $\sqrt s_{NN} \ge 7.6$ GeV \cite{KAB:89a}. Therefore, in order to correctly interpret the high energy quasi-plateaus here  we
use the results of  meta-analysis \cite{Metaanalisys} of the quality of  data description  (QDD) of 10 existing event generators of nucleus-nucleus collisions along with the thorough analysis of  irregularities in the existing experimental hadron yield ratios. 
 
The work is organized as follows.  In Sect. 2 we remind the basic elements of the HRGM with multicomponent hard-core repulsion. A brief  description of the meta-analysis  suggested in \cite{Metaanalisys} is presented in Sect. 3 along with a discussion of existing hadron multiplicity  data which  help to shed  light on the problem of the formation of two quark-gluon-hadron mixed phases.  In Sect. 4  our conclusions are formulated.

\section{ HRGM with multicomponent hard-core repulsion}
\label{sect2_Bugaev}

The HRGM is based on the assumption of  local thermal and chemical equilibrium at CFO. 
Hence the hadron yields  produced in the collisions of large atomic nuclei can be found using
the grand canonical valuables, i.e.  using the temperature $T$, the baryonic $\mu_B$, the strange $\mu_s$ and the isospin third projection $\mu_{I3}$ chemical potentials.  As usual, the chemical potential $\mu_s$ is fixed by the condition of zero total strange charge. The possible deviation of strange charge from the full chemical equilibrium  is taken into account by the parameter $\gamma_s$ \cite{Rafelski}. 
It changes the  thermal density $\varphi_j$ of hadron sort $j$  as  
 $\varphi_j\rightarrow\gamma_s^{S_j}\varphi_j$,
where $S_j$ is the total number of strange valence quarks and antiquarks in such a hadron. 

\begin{figure}[h]
\centering
\mbox{\includegraphics[width=80mm,clip]{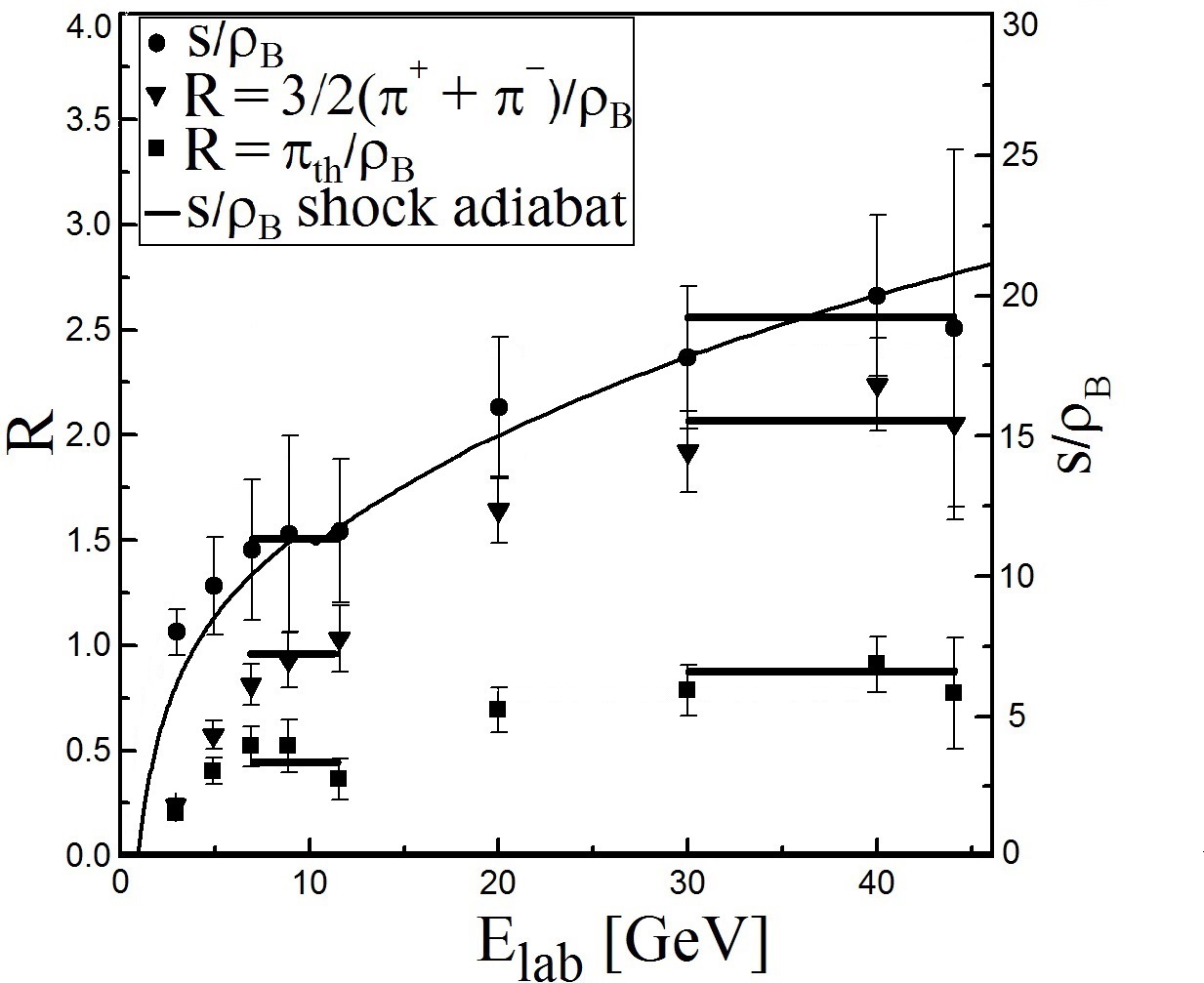}\hspace*{-4.4mm}\includegraphics[width=65mm,clip]{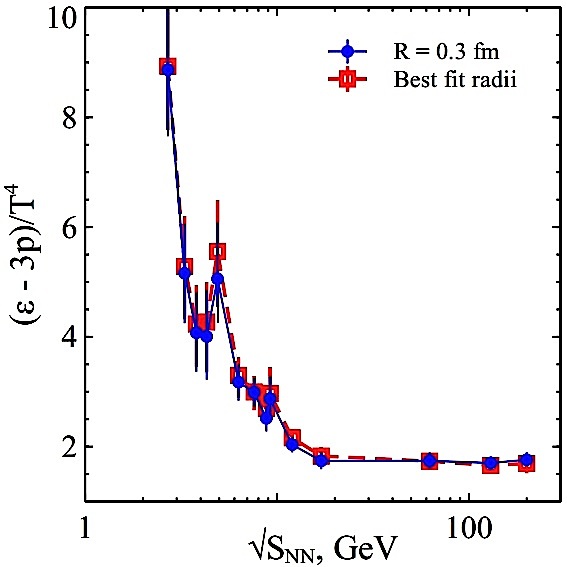}}
\caption{Left: The correlated quasi-plateaus at CFO  found in \cite{Bugaev:SA1} (see details in the text). 
Right: Trace anomaly as function of collision energy at CFO  established  in \cite{Bugaev:SA2}. }
\label{fig_Bugaev_plateaus}       
\end{figure}

The main difference of the present version of HRGM from the  ones developed by other authors  is that in our HRGM 
several sorts of hadrons  have  individual hard-core radii. Thus, it employes 
different  hard-core radii for pions, $R_{\pi}$, kaons, $R_K$,  $\Lambda$-hyperons, $R_\Lambda$, 
other mesons, $R_m$, and baryons, $R_b$. 
The best global fit of  111 independent hadronic multiplicities measured 
in the whole collision energy range from  $\sqrt{s_{NN}} =2.7$
GeV to $\sqrt{s_{NN}} = 200$ GeV
was found for $R_b$ = 0.355 fm, $R_m$ = 0.4 fm, $R_{\pi}$ = 0.1 fm,  $R_K$ = 0.37 fm   
and $R_\Lambda = 0.11$ fm with  the quality  $\chi^2/dof \simeq 0.95$ \cite{Sagun,Sagun2}.
The second virial coefficient  between the hadrons  of  hard-core radii $R_i$ and $R_j$ is defined  as $b_{ij}=\frac{2\pi}{3}(R_i+R_j)^3$.
Taking from  the thermodynamic code THERMUS \cite{Thermus} such  characteristics of hadrons of sort $i$ 
as the spin-isospin degeneracy $g_i$, the mass $m_i$ and the width $\Gamma_i$ one can find  the set of partial pressures
$p_i$ for each hadronic 
component ($p=\sum_i p_i$ is total pressure) from the system
\begin{equation}
\label{EqI}
p_i=T\varphi_i\exp\left[\frac{\mu_i-2\sum_j p_jb_{ji}+\sum_{jl}p_j b_{jl}p_l/p}{T}\right] \,.
\end{equation}
Here $\mu_i=Q^B_i\mu_B+Q^{I3}_i\mu_{I3}+Q^S_i\mu_S$ is the full chemical potential of the $i$-th hadronic sort expressed via its   charges $\{Q^A_i\}$  and the corresponding chemical potentials $\{\mu_A\}$ (with $A \in \{B, I3, S \}$).  
In the Boltzmann approximation  the thermal density of $i$-th hadronic  sort reads as 
\begin{equation}
\label{EqII}
\varphi_i=\gamma_s^{S_i}g_i \hspace*{0.0cm}\int\limits_{M_i}^\infty \hspace*{0.cm}dm \, f(m,m_i,\Gamma_i) \hspace*{-0.cm}
\int \hspace*{-0.cm}
\frac{k^2 d k}{2\pi^2} \hspace*{0.1cm}e^{- \frac{\sqrt{m^2+{k}^2} }{T} } \,,
\end{equation}
where
$M_i$ is  a threshold  of its dominant decay channel and $f$ is the normalized  Breit-Wigner mass attenuation function. 
Thermal multiplicities $N_i^{th}=V\frac{\partial p}{\partial\mu_i}$ ($V$ is the effective  volume at CFO) should be 
corrected by the hadron decays after the CFO according to the branching ratios $Br_{l\rightarrow i}$.  The latter  define 
the probability of particle $l$ to decay into a particle $i$. Hence  the ratio of full  multiplicities can be written  as
\begin{equation}
\label{EqIII}
\frac{N^{tot}_i}{N^{tot}_j}=
\frac{p_i+\sum_{l\neq i}p_lBr_{l\rightarrow i}}{p_j+\sum_{l\neq j}p_lBr_{l\rightarrow j}}\,.
\end{equation}
With the help of  (\ref{EqIII}) we obtained the high quality fit of  experimental hadron ratios measured at AGS for energies  $\sqrt s_{NN}=2.7, 3.1, 3.6, 4.3, 4.9$ GeV\cite{AGS1,AGS2,AGS2b,AGS_p2,AGS3,AGS4,AGS5,AGS6,AGS7,AGS8},  at SPS energies  $\sqrt s_{NN}=6.3, 7.6, 8.8, 12.3, 17.3$ GeV  measured  by the NA49 Collaboration  \cite{SPS1,SPS2,SPS3,SPS4,SPS5,SPS6,SPS7,SPS8,SPS9} and   at 
RHIC energies  $\sqrt s_{NN}=9.2, 62.4, 130, 200$ GeV measured  by the STAR Collaboration \cite{RHIC}.
As described in \cite{SFO:13,HRGM:13}, from these data we constructed 111 independent ratios measured at 14 values of collision energies.  The most important results are shown in Figs.~\ref{fig_Bugaev_Horn} and \ref{fig_Bugaev_plateaus}.
The left panel of Fig.~\ref{fig_Bugaev_plateaus} shows 
the highly correlated quasi-plateaus in the collision energy dependence of the entropy per baryon $s/\rho_B$, total pion number per baryon $\rho_{\pi}^{tot}/\rho_B$, and thermal pion number per baryon $\rho_{\pi}^{th}/\rho_B$ at laboratory energies 6.9--11.6 GeV (i.e. $\sqrt{s_{NN}} = 3.6-4.9$ GeV) which were  found in \cite{Bugaev:SA1}.   As one can see from the left panel of   Fig.~\ref{fig_Bugaev_plateaus},  a clear plateau is demonstrated by  the thermal pion number per baryon 
while other quantities show quasi-plateaus. Nevertheless,  all these quasi-plateaus  are important, since their strong 
correlation with the plateau  in  the thermal pion number per baryon
allows one  to find out  their common width in  the collision  energy \cite{Bugaev:SA1,Bugaev:SA2}.

Note that  these low energy  quasi-plateaus were predicted about 25 years  ago \cite{KAB:89a,KAB:90,KAB:91} 
as a manifestation of  the  anomalous thermodynamic properties of  quark-gluon-hadron mixed phase. 
In contrast to the normal thermodynamic properties, in the medium with the anomalous  ones  the adiabatic compressibility of matter increases for  increasing pressure. In the normal media (pure gaseous or liquid phases) there exists a repulsion between the constituents at short distances which leads to an opposite behavior of the adiabatic compressibility. Therefore, an appearance
of these quasi-plateaus is a signal of  the  quark-gluon-hadron mixed phase formation \cite{KAB:89a,KAB:90,KAB:91}. Such a conclusion is strongly supported by an existence of  the sharp peak of the trace anomaly $\delta = 
\frac{\varepsilon - 3p}{T^4}$ (here $\varepsilon$ is energy density) at $\sqrt{s_{NN}} = 4.9$ GeV  \cite{Bugaev:SA2}
(see the right panel of Fig.~\ref{fig_Bugaev_plateaus}).  This peak is important, since in lattice QCD  
an inflection or a maximum point of the trace anomaly is used for a determination of the pseudo-critical temperature of the cross-over  transition \cite{lQCD}.  One may think that  a sharp peak of $\delta$ at CFO   is  exclusively generated by  the peak of baryonic density which in our HRGM also exists at $\sqrt{s_{NN}} = 4.9$ GeV.  However, the real situation is more complicated.  Writing the trace anomaly as 
\begin{eqnarray}\label{EqIV}
\delta =  \frac{\varepsilon - 3p}{T^4} = \frac{Ts + \mu_B \rho_B  + \mu_{I3} \rho_{I3} - 4 p}{T^4} \simeq  \frac{s}{T^3} \left( 1 + \frac{\mu_B}{T} \frac{\rho_B}{s} \right)  - 4 \frac{ p}{T^4} \,,
\end{eqnarray}
where in the last step the small contribution $\mu_{I3} \rho_{I3}$  related to the charge of the third isospin projection is neglected. From (\ref{EqIV}) one can easily conclude that  the strong increase of $\delta$ on the collision energy
 interval   $\sqrt{s_{NN}} = [4.3; 4.9]$ GeV   is  provided by a strong jump of the effective number of degrees of freedom 
 ${s}/{T^3}$ on this interval \cite{Bugaev:SA2,NonsmFO}.  Note that despite  the existence of a baryon density peak on this interval of collision energy,   the entropy per baryon ${s}/{\rho_B}$ is constant on it  as one can see from the left panel of  Fig.~\ref{fig_Bugaev_plateaus}. Now it is evident that without  a strong jump of the effective number of degrees of freedom 
 ${s}/{T^3}$ the sharp peak of trace anomaly at CFO would not exist. At higher  collision energies the trace anomaly $\delta$  decreases mainly because the ratio $\mu_B/T$ strongly decreases, while  the inverse entropy per baryon decreases  slowly. 
 It is important to mention  that the sharp peak of  $\delta$ is seen, if    the finite width of all hadronic resonances  is included into the HRGM \cite{NonsmFO}, while for the HRGM with a zero width of hadron resonances such a peak is washed out. 

The physical origin of the trace anomaly sharp peak (and, hence, of a strong jump of the effective number of degrees of freedom ${s}/{T^3}$)  at CFO found at  $\sqrt{s_{NN}} = 4.9$ GeV is rooted in the trace anomaly peak  existing  at the shock adiabat \cite{Bugaev:SA2}.  
The shock adiabat model reasonably well   describes the hydrodynamic and thermodynamic  quantities  of the initial state formed in the central nucleus-nucleus collisions in the laboratory  energy range $1$ GeV $ \le E_{lab} \le $ 30 GeV \cite{Bugaev:SA2}, while at higher collision energies, i.e. for  $\sqrt{s_{NN}} \ge 7.6$ GeV,  it can be used for qualitative estimates. 
In  \cite{Bugaev:SA2} it was found  that the peak of $\delta$ at the shock adiabat appears at the collision energy corresponding exactly to the boundary between the quark gluon plasma (QGP) and quark-gluon-hadron mixed phase and, therefore, the trace anomaly sharp peak at CFO is a signal of QGP formation.  
In this respect it is interesting  that 
in the right panel of  Fig.~\ref{fig_Bugaev_plateaus} one can see a second peak of trace anomaly located at $\sqrt{s_{NN}} = 9.2$ GeV.  Although the second peak of $\delta$ is less pronounced than the first one, it is also associated to the high energy set of quasi-plateaus 
shown in the left panel of Fig.~\ref{fig_Bugaev_plateaus}  at the collision energy interval $E_{lab} = [30; 44.]$ GeV
($\sqrt{s_{NN}} = [7.6; 9.2]$ GeV). Therefore,  the future experiments at RHIC,  NICA and FAIR will have to find out whether the high energy peak of trace anomaly has any physical meaning.

\section{Meta-analysis of quality of data description}

The main objects of  a meta-analysis suggested in \cite{Metaanalisys} are the mean deviation squared  of the quantity $A^{model,h}$ of the model  M from the data $A^{data,h}$ per number of  the data points $n_d$ for  a particle type $h$  
 \begin{eqnarray}\label{EqV}
 \langle\chi^2/n\rangle^h_A\biggl|_{M} = \frac{1}{n_d} \sum\limits_{k=1}^{n_d} \left[ \frac{A^{data,h}_k -A^{model,h}_k }{\delta A_k^{data,h}}\right]^2  \biggl|_{M}  \,,
\end{eqnarray}
and its error which is defined according to the rule of indirect measurements \cite{Taylor82} as 
 \begin{eqnarray}\label{EqVI}
 \Delta_{A}\langle\chi^2/n\rangle^h_A\biggl|_{M} &\equiv & \left[ \sum\limits_{k=1}^{n_d}  \left[\delta A_k^{data,h} \,  \frac{\partial \langle\chi^2/n\rangle^h_A\biggl|_{M}}{\partial ~A^{data,h}_k} \right]^2  \right]^\frac{1}{2}   
 \equiv \frac{2}{\sqrt{n_d}} \sqrt{\langle\chi^2/n\rangle^h_A\biggl|_{M}} \,,~
\end{eqnarray}
where $\delta A_k^{data,h}$ is an experimental error of the experimental quantity $A^{data,h}_k$ and the summation in Eqs. (\ref{EqV}) and (\ref{EqVI})  runs over all data point $n_d$ at given collision energy.  
For a convenience the quantity defined in (\ref{EqV})  is called the quality of data description (QDD). 
To get  the most complete  picture of the  dynamics of nuclear collisions, one has to compare the available data on 
  the transverse mass ($m_T$) distributions $A= \frac{1}{m_T}\,\frac{d^2 N(m_T, y)}{d m_T dy}$, the longitudinal rapidity ($y$) distributions $A=\frac{d N(y)}{dy}$ and the hadronic yields (Y) measured  at  midrapidity $A=\frac{d N(y=0)}{dy}$  or/and the total one, i.e. measured within   4$\pi$ solid angle, 
since right  these observables are traditionally believed to be sensitive to the equation of state properties \cite{Rafelski2,Shuryak}. 
 
 The QDD of strange hadrons was found for  two types of models \cite{Metaanalisys}:
\begin{itemize}

\item {\bf The  hadron gas (HG) models} are as follows: ARC \cite{Arc},  RQMD2.1(2.3) \cite{RQMD}, HSD \cite{HSDgen,HSDgen2,S5.4a},  UrQMD1.3(2.0, 2.1, 2.3) \cite{Bass98}, statistical hadronization model (SHM) \cite{SHMgen} and AGSHIJET\_N* \cite{HIJETa,S5.4b}.  These models  do not include the QGP formation in the process of  A+A collisions.

\item {\bf The  QGP models} are as follows:  Quark Combination  (QuarkComb) model \cite{QComb}, 3-fluid dynamics (3FD) model \cite{3FDgenA,3FDgen,3FDII,3FDIII},  PHSD model \cite{PHSDgen,PHSDgen2,Phsd} and Core-Corona model \cite{CoreCorMa,CoreCorMb}.   
These  generators   explicitly assume the QGP formation in A+A collisions.
\end{itemize}
{A short description of these models along with the criteria of their selection can be found in the Appendix of \cite{Metaanalisys}.}

The main idea of such a meta-analysis \cite{Metaanalisys} is based on the assumption  that  the HG models of  heavy ion  collisions should provide worse description of the  data above  the  QGP threshold energy,  whereas below this threshold they should be able to better (or at least not worse) reproduce experimental data compared to the QGP models. 
Furthermore, it is assumed  that  both kinds of models should provide an equal  and rather  good QDD at the 
energy  of mixed phase production. Hence,  the energy of the mixed phase formation  should be  located below the energy 
at which the equal QDD  is changed to the essentially  worse   QDD  of HG  models.
 
\begin{figure}[h]
\centering
\includegraphics[width=140mm,clip]{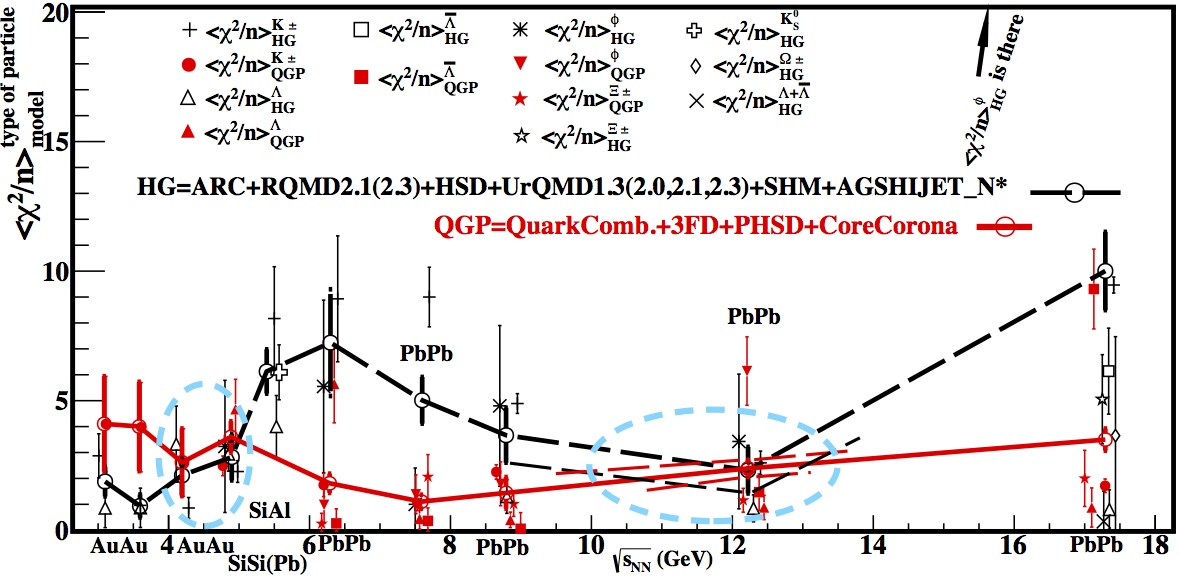}
\caption{Comparison of  $\langle\chi^2/n\rangle^{\overline{\{h\}}}_{\overline{\{A\}}}\biggl|_{\overline{HG}}$
 (black symbols and dashed curve) and 
$\langle\chi^2/n\rangle^{\overline{\{h\}}}_{\overline{\{A\}}}\biggl|_{\overline{QGP}}$  (red symbols and solid curve) as functions of collision energy obtained for the arithmetic averaging. The symbols of different hadrons which  correspond to the same collision energy are slightly spread around the energy value for better perception. 
The symbols are connected by the lines to guide the eye. The numbers staying behind the short name of the model indicate the version(s) used in meta-analysis. The short dashed ovals indicate the regions of possible mixed phase formation.}
\label{fig_Bugaev_Chi}       
\end{figure}

Based on these assumptions the experimental data measured at the collision energies $\sqrt{s_{NN}}\,=$   3.1, 3.6, 4.2, 4.9, 5.4, 6.3, 7.6, 8.8,   12.3 and 17.3 GeV were analyzed in \cite{Metaanalisys}.   The collision energies  $\sqrt{s_{NN}}\,\le $  4.9 GeV correspond to  Au+Au  reactions studied at AGS. At  $\sqrt{s_{NN}}\,=$ 5.4 GeV the reactions Pb+Si, Si+Si and 
Si+Al were also  investigated   at AGS,  while higher collision energies  correspond to Pb+Pb reactions studied at SPS. 
Using the definitions (\ref{EqV}) and  (\ref{EqVI})  at each collision energy the available description of  the   transverse mass  distributions,  the longitudinal rapidity  distributions and the hadronic yields  measured  at  midrapidity or/and the total one obtained by a given model  was  arithmetically averaged for each kind of analyzed  strange hadron.  The QDDs  and their errors  obtained   for each energy and for each hadron were  arithmetically averaged over  the models belonging to the same type. 
Then the QDDs  and their errors found in this way for the models belonging to the same type  were arithmetically averaged 
for  each hadron and  antihadron, if available,   in order to reduce the number of data for a comparison.  
Finally, the resulting QDDs  and their errors of the same same type of model   were arithmetically averaged over all hadronic species. More details can be found in \cite{Metaanalisys}. The averaged QDDs  
 of HG models  $\langle\chi^2/n\rangle^{\overline{\{h\}}}_{\overline{\{A\}}}\biggl|_{\overline{HG}}$  and  the ones of  QGP models $\langle\chi^2/n\rangle^{\overline{\{h\}}}_{\overline{\{A\}}}\biggl|_{\overline{QGP}}$  were found in this way  together with their errors.  The results are shown in Fig.~\ref{fig_Bugaev_Chi}.
 
 From Fig.~\ref{fig_Bugaev_Chi} one can see that the meta-analysis of  work
\cite{Metaanalisys}  leads to an independent  conclusion that the mixed phase exists at the same collision energy range 
$\sqrt{s_{NN}} = [4. 3; 4.9]$ GeV
which  was originally  found in \cite{Bugaev:SA1,Bugaev:SA2}.
This result is important not only to  validate  the entire framework of shock adiabat model used in \cite{Bugaev:SA1,Bugaev:SA2}, but also to justify  the jump of the effective number of degrees of freedom $s/T^3$ at CFO  and a sharp peak of the trace anomaly $\delta$ at CFO as reliable signals of  QGP formation.
In addition the meta-analysis of QDD \cite{Metaanalisys} predicts that the most probable  collision energy range of the second phase transition is $\sqrt{s_{NN}} =10-13.5$ GeV.  Thus, the meta-analysis of QDD  supports  an interpretation of  the
 high energy set of correlated quasi-plateaus   as an indicator of phase transition, although it  shifts  this transition to slightly higher collision energies. 
 
Unfortunately, at present it is  impossible to distinguish between two possible explanations of this phenomenon.
The first possible explanation is that with increasing collision energy the initial states of  thermally equilibrated matter formed in nucleus-nucleus collisions move first  from the hadron gas into the mixed phase, then from the mixed phase to QGP  and then again they return to the same mixed phase, but at higher initial temperature and lower 
baryonic density. 
Such a scenario corresponds to the case, if QCD  phase digram has a critical endpoint \cite{Metaanalisys}. An alternative explanation \cite{Metaanalisys} corresponds to the QCD phase diagram with the tricritical endpoint. In such a case the second phase transition is the second order phase  transition of (partial) chiral symmetry restoration  or  a transition between quarkyonic matter and QGP \cite{QYON}.  It is necessary to stress,  that despite the lack of  a single interpretation of the second phase transition  possibly   existing at $\sqrt{s_{NN}} =10-13.5$ GeV there are strong arguments \cite{Metaanalisys} that the (tri)critical endpoint of  QCD phase diagram maybe located at $\sqrt{s_{NN}} =12-14$ GeV.

\begin{figure}[h]
\centering
\mbox{\includegraphics[width=75mm,clip]{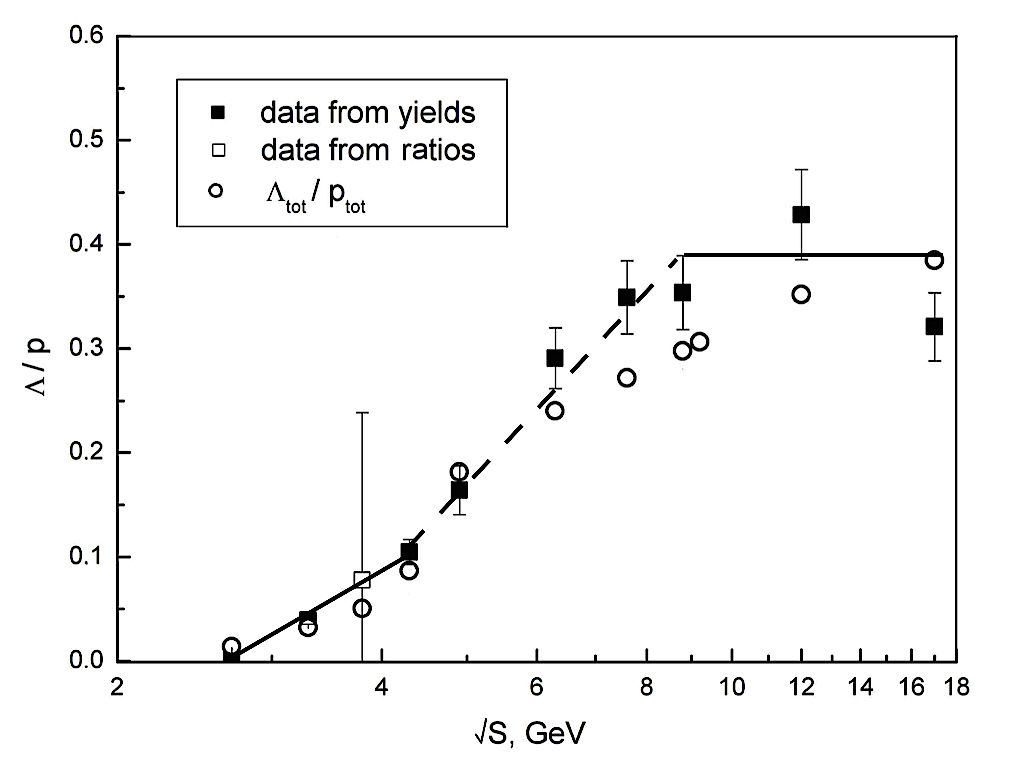}\hspace*{-4.2mm}\includegraphics[width=72mm,clip]{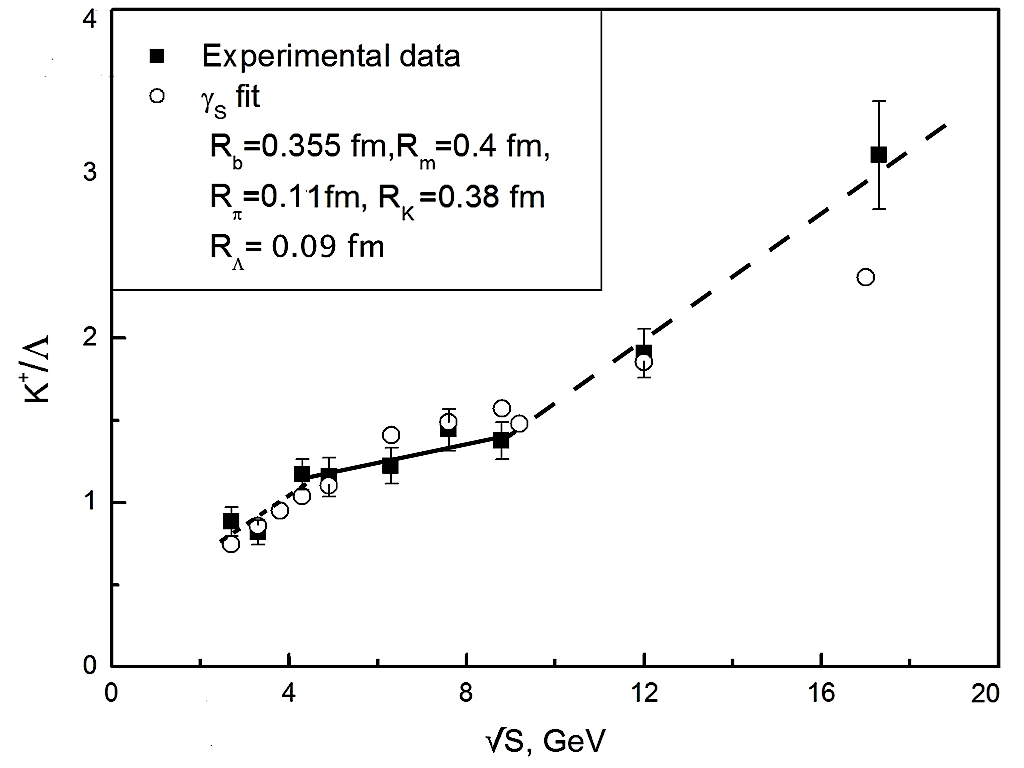}}
\caption{The collision energy dependence of the $\Lambda/p$  (left) and   $K^+/\Lambda$ (right) ratios obtained within the present HRGM. The lines are given to guide the eye. More explanations are given in the text. }
\label{fig_Bugaev_Lamd}       
\end{figure}

The combined conclusions obtained from inspecting two sets of correlated quasi-plateaus at CFO,  two peaks of trace anomaly at CFO and 
the ones found out by the meta-analysis, led us  a thorough analysis of  the   $\Lambda/p$ and  $K^+/\Lambda$ ratios. 
From the left panel of Fig.~\ref{fig_Bugaev_Lamd} one can see  that there are three regimes in the
energy dependence of  the $\Lambda/p$
ratio: at  $\sqrt{s_{NN}}=4.3$ GeV the slope of this ratio clearly increases, while above   
$\sqrt{s_{NN}}=8.8$ GeV it nearly saturates.  The sudden  increase  of  $\Lambda/p$ slope  
at  $\sqrt{s_{NN}}=4.3$ GeV can be 
naturally explained by  the idea  of  work  \cite{Rafelski:82} that the mixed phase formation can be 
identified by a rapid increase in the number of strange quarks per light quarks.  
Evidently, the   $\Lambda/p$    ratio is a  convenient indicator because at low collision energies $\Lambda$ hyperons
are generated in collisions of nucleons.  Moreover, such a ratio does not depend on baryonic chemical potential, since both the protons and $\Lambda$ hyperons have the same baryonic charge. 
As it is seen from the left panel of  Fig.~\ref{fig_Bugaev_Lamd}, this mechanism works up to  $\sqrt{s_{NN}}=4.3$ GeV,
while an appearance of the mixed phase should lead to an  increase 
of  the number of strange quarks and antiquarks  due to the annihilation of light quark-antiquark and gluon pairs.
Clearly, this simple  picture is well fitted into the prediction that the mixed phase can be reached at  $\sqrt{s_{NN}}=4.3$ GeV,
while QGP is formed at $\sqrt{s_{NN}} \ge 4.9$ GeV.  The dramatic decrease of the slope  of the experimental  $\Lambda/p$
ratio  at $\sqrt{s_{NN}} > 8.8$ GeV which is seen in  Fig.~\ref{fig_Bugaev_Lamd} can be an evidence for the second 
phase transformation, which we discussed above.  

It is appropriate  to say 
a few words about the experimental data of hadron yields shown in both panels of Fig.~\ref{fig_Bugaev_Lamd}.
For the AGS collision energies $\sqrt{s_{NN}}=$ 2.7, 3.1, and 4.3 GeV the yields of 
protons and kaons were, respectively,  taken from  Refs.\ \cite{AGS8} and \cite{AGS_p2},   whereas  for  $\Lambda-$hyperons  they were taken from  Ref. \cite{AGS3}. 
Experimental yields measured at the highest AGS energy  $\sqrt{s_{NN}}=$
4.9 GeV for protons and kaon were taken from Ref.\  \cite{AGS2,AGS8}, while  
for $\Lambda$ they were given in Ref.\ \cite{AGS7}.  
The mid-rapidity yields of protons, kaons  and lambdas 
measured  at the SPS energies $\sqrt{s_{NN}}=$
6.3, 7.6, 8.8, 12, and 17.3 GeV are provided  by the NA49 collaboration in Refs.\
\cite{SPS2,SPS3,SPS5,SPS6,SPS7}.
 For a comparison, in Fig.~\ref{fig_Bugaev_Lamd}  we also show the value with huge error bars which are  found from 
other two ratios, $\Lambda/\pi^-$ and $p/\pi^-$, for  $\sqrt{s_{NN}}=3.6$ GeV.

It is interesting that the energy dependence of the $K^+/\Lambda$ ratio shows the change of slopes at the same energies
as the $\Lambda/p$ ratio. This  can be  see from the right panel of  Fig.~\ref{fig_Bugaev_Lamd}.  Note that  in the dominant  hadronic reactions the positive
kaons and $\Lambda$ hyperons are born simultaneously.  Since both of these hadrons carry the strange charge, then the logic of 
work  \cite{Rafelski:82} is inapplicable to their ratio.  Therefore, in contrast to an increase of the slope of the $\Lambda/p$ ratio on the interval  $\sqrt{s_{NN}}= [4.3; 8.8]$ GeV, the $K^+/\Lambda$ ratio  has a flattening  of the slope on this collision energy interval. In the leading order  this   ratio is  defined  via the kaon mass $m_K$,  the $\Lambda$ mass $m_\Lambda$ and two chemical potentials as $K^+/\Lambda \simeq \sqrt{ \left[ \frac{m_K}{m_\Lambda}\right]^3}\exp\left[ \frac{m_\Lambda - m_K + 2\mu_S - \mu_B}{T} \right]$. Therefore, a small slope of this ratio at the interval $\sqrt{s_{NN}}= [4.3; 8.8]$ GeV evidences about the cancellation of energy dependencies of  strange  and  baryonic chemical potentials, i.e. $m_\Lambda - m_K + 2\mu_S - \mu_B \simeq Const$ for $\sqrt{s_{NN}}= [4.3; 8.8]$ GeV. An increase of the $K^+/\Lambda$ ratio at higher collision energies can be mainly explained by the fast decrease of the baryonic chemical potential.

\begin{figure}[h]
\centering
\sidecaption
\includegraphics[width=84mm,clip]{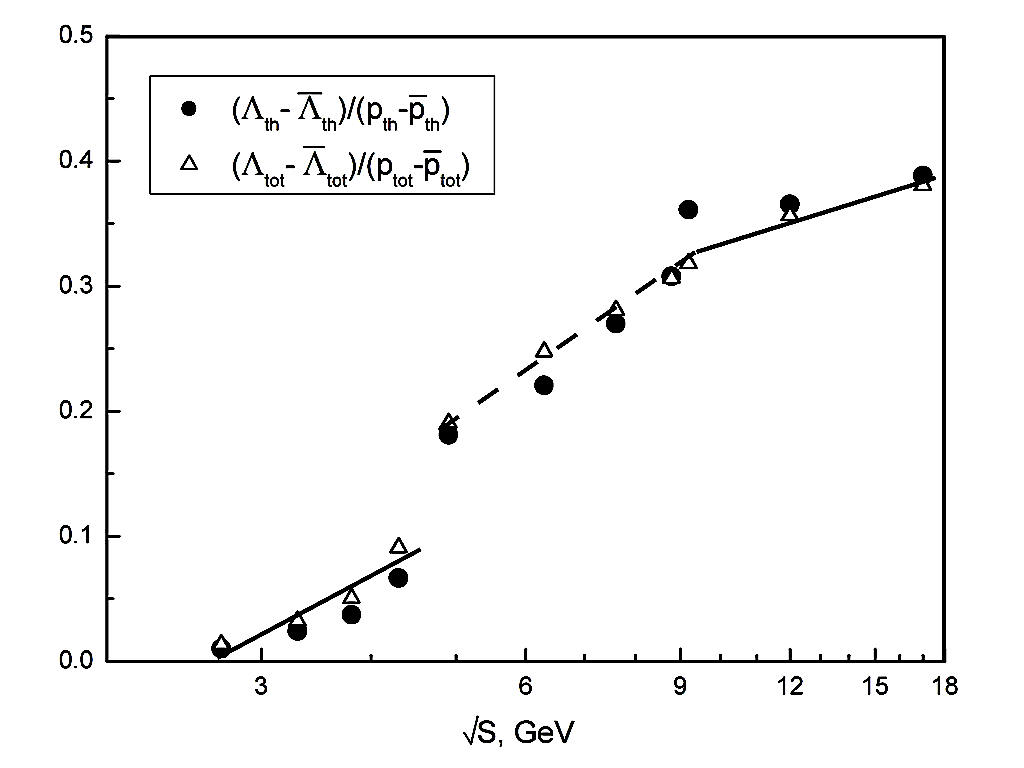}
\caption{Most recent predictions for the collision energy dependence of the  $\frac{\Delta \Lambda}{\Delta\, p}$ ratio. The triangles depict the ratio of total multiplicities, while the circles correspond to the ratio of  thermal multiplicities. The lines are given to guide the eye. }
\label{fig_Bugaev_DLam}       
\end{figure}

Very recently a better description of  the $\Lambda/p$ ratio was achieved in \cite{Bugaev15a}, when more data were analyzed. 
Note that this result was obtained not on the expense of worsening of other hadron yield ratios. Based on this new fit of hadron yield ratios the predictions for the  $\frac{\Delta \Lambda}{\Delta\, p} =  \frac{\Lambda - \bar \Lambda}{p - \bar p}$ ratio are made. As one can see from Fig.~\ref{fig_Bugaev_DLam} this ratio demonstrates even more dramatic changes in the collision energy dependence.
Indeed,  at the narrow  collision energy interval  $\sqrt{s_{NN}} =4.3-4.9$ GeV this ratio  has  a strong jump, while 
at $\sqrt{s_{NN}} = 9.2$ GeV it  shows a change of  slope. 
 Our educated guess  is that the collision energy dependence  of  the  $\frac{\Delta \Lambda}{\Delta\, p}$  ratio  is an indicator of two phase transformations \cite{Bugaev15a}. Since the observed jump of this ratio  is located in  the collision energy region of the mixed phase formation (i.e. with a first order phase transition), then a change of  its   slope at $\sqrt{s_{NN}} = 9.2$ GeV  can be naturally associated with a weak first order or a second order phase transition.  Note that such a hypothesis is well supported by  the results of the meta-analysis \cite{Metaanalisys} which is briefly  summarized above.

\section{Conclusions}

From the discussions given  in  previous sections it is clear that a development of the multicomponent version of HRGM in 2012 led to a real breakthrough in our understanding of the thermodynamics at CFO. With the help of HRGM it was possible for the first time 
to describe the Strangeness Horn with the  highest accuracy \cite{HRGM:13,SFO:13,Sagun2} including the topmost point. The new concept of separated CFOs for strange and non-strange  hadrons with conservation laws connecting them  allows one to naturally  explain the appearance of apparent chemical non-equilibrium of strange particles \cite{SFO:13}.  Furthermore, a thorough analysis of the entropy per baryon and the pion number (thermal and total) per baryon at CFO led to a finding out of two sets of strongly correlated quasi-plateaus \cite{Bugaev:SA1,Bugaev:SA2}. Since the low energy set of  quasi-plateaus was predicted in \cite{KAB:89a,KAB:90, KAB:91} as a signal of the quark-gluon-hadron mixed phase formation, then it was necessary to give a physical interpretation of the high energy set of  quasi-plateaus.   

A good hint to interpret the appearance of high energy set of quasi-plateaus is provided by the meta-analysis \cite{Metaanalisys} of  QDD. Since the QDD meta-analysis gave an independent evidence for the quark-gluon-hadron mixed phase formation at the narrow region of collision energy $\sqrt{s_{NN}} = 4.3-4.9$ GeV, then its predictions for the 
possible existence of another  mixed phase at collision energies $\sqrt{s_{NN}} = 10-13.5$ GeV led us to  a more thorough inspection of existing hadron multiplicity ratios.   As one can see from  the left panel of Fig.~\ref{fig_Bugaev_Lamd} 
the $\Lambda/p$ ratio exhibits  three different regimes in the collision energy dependence: at  $\sqrt{s_{NN}}=4.3$ GeV the slope of this ratio suddenly  increases, while above  $\sqrt{s_{NN}}=8.8$ GeV it nearly saturates. As we argued above 
a  strong  increase  of  $\Lambda/p$ slope at  $\sqrt{s_{NN}}=4.3$ GeV can be 
naturally explained by  the idea  of  work  \cite{Rafelski:82} that the mixed phase formation can be 
identified by a rapid increase in the number of strange quarks per light quarks, while a dramatic decrease of  the $\Lambda/p$ slope at $\sqrt{s_{NN}} > 8.8$ GeV  can be an evidence  for the second phase transformation. 

It is remarkable that the  $K^+/\Lambda$ ratio   shows the change of slopes at the same energies
as the $\Lambda/p$ ratio, although the  $K^+/\Lambda$ ratio involves two strange particles and (in general) it strongly depends on the baryonic chemical potential. Also the trace anomaly at CFO  shows two peaks  at the collision energies $\sqrt{s_{NN}}= 4.9$ GeV and
$\sqrt{s_{NN}}= 9.2$ GeV. The low energy trace anomaly peak can be explained within the shock adiabat model   \cite{Bugaev:SA2} as a signal of QGP formation, whereas the existence of  high energy peak requires further confirmation by  better experimental data.  If it will be confirmed, then  it will serve as a new indicator for a second phase transition. 
Nevertheless, already now all  irregularities at CFO  discussed here together with the results of the QDD meta-analysis form a coherent picture of  possible observation of two mixed phases in nucleus-nucleus collisions. 
Therefore,  the Beam Energy Scan program at RHIC has a unique chance to experimentally verify the above signals and to discover the mixed phases before  the start of NICA and FAIR programs. The new observable, the  $\frac{\Delta \Lambda}{\Delta\, p}$  ratio suggested in  \cite{Bugaev15a}, will be very helpful for this because of its high sensitivity.    However, to reach such goals the RHIC experiments should provide much smaller error bars, especially at low energies. Hence, the experiments in a fixed target mode are absolutely necessary for the success of  the Beam Energy Scan program at RHIC.

\begin{acknowledgement} The authors are thankful to D. B. Blaschke, T. Galayuk, R. A. Lacey, I. N. Mishustin, D. H. Rischke, K. Redlich, L. M. Satarov,  A.V. Taranenko, K. Urbanowski and Nu Xu for interesting  discussions and valuable comments. K.A.B., A.I.I., V.V.S. and G.M.Z. acknowledge the  partial support of the program ``On perspective fundamental research in high-energy and nuclear physics'' launched by the Section of Nuclear Physics of NAS of Ukraine. D.R.O. acknowledges a support of Deutsche Telekom Stiftung.  K.A.B. is very thankful to all  organizers of the ICNFP2015 for providing him with  a chance to present and to discuss these results at the Conference and for a warm hospitality in OAC. 
\end{acknowledgement}

%
%
%

\end{document}